\documentclass[sigconf]{acmart}

\renewcommand\footnotetextcopyrightpermission[1]{} 
\AtBeginDocument{%
  }

\setcopyright{acmlicensed}
\copyrightyear{2018}
\acmYear{2018}
\acmDOI{XXXXXXX.XXXXXXX}

\acmConference[Conference acronym 'XX]{Make sure to enter the correct
  conference title from your rights confirmation emai}{June 03--05,
  2018}{Woodstock, NY}

\acmISBN{978-1-4503-XXXX-X/18/06}

\usepackage{url}
\usepackage[capitalise]{cleveref}
\usepackage[ruled,linesnumbered]{algorithm2e}
\usepackage{tikz}
\usepackage{amsmath}
\usepackage{microtype}

\usepackage{listings, xcolor}
\usepackage[ruled]{algorithm2e} %
\usepackage{ragged2e}

\usepackage{array}
\usepackage{multirow}

\usepackage{tcolorbox}
\tcbset{boxrule = 0.1mm,colback=white!5!white}

\usepackage{subcaption}
\usepackage{booktabs}
\usepackage{pbalance}
\usepackage[normalem]{ulem}

\usepackage{pifont}
\newcommand{\para}[1]{\vspace{2pt}\noindent\textbf{#1.~}}
\newcommand{\ignore}[1]{}
\newcommand{\system}{\textit{\sloppy{Poker}\@}}
\newcommand{\Pattern}{\textit{\sloppy{Sneaky}\@}}
\newcommand{\PowPermiss}{\textit{\sloppy{Permisson-Related}\@}}
\newcommand{\PowNecess}{\textit{\sloppy{Permisson-Unrelated}\@}}
\newcommand{\PowProm}{\textit{\sloppy{Promotional}\@}}
\newcommand{\PowFun}{\textit{\sloppy{Functional}\@}}

\begin{document}

\title{Understanding the Sneaky Patterns of Pop-up Windows in the Mobile Ecosystem}


\author{Dongpeng Wu}
\affiliation{%
  \institution{Sun Yat-sen University}
  \country{}
}
\email{wudp8@mail2.sysu.edu.cn}

\author{Yuhong Nan}
\authornote{Corresponding author: Yuhong Nan}
\affiliation{%
 \institution{Sun Yat-sen University}
   \country{}
}
\email{nanyh@mail.sysu.edu.cn}

\author{Shaojiang Wang}
\affiliation{%
  \institution{Sun Yat-sen University}
    \country{}
}
\email{wangshj53@mail2.sysu.edu.cn}

\author{Jiawei Wang}
\affiliation{%
  \institution{Sun Yat-sen University}
    \country{}
}
\email{wangjw99@mail2.sysu.edu.cn}

\author{Luwa Li}
\affiliation{%
  \institution{Sun Yat-sen University}
    \country{}
}
\email{lilw33@mail2.sysu.edu.cn}

\author{Xueqiang Wang}
\affiliation{%
  \institution{University of Central Florida}
    \country{}
}
\email{xueqiang.wang@ucf.edu}

%


\begin{abstract}
In mobile applications, Pop-up window (PoW) plays a crucial role in improving user experience, guiding user actions, and delivering key information. Unfortunately, the excessive use of PoWs severely degrades the user experience. These PoWs often sneakily mislead users in their choices, employing tactics that subtly manipulate decision-making processes.
In this paper, we provide the first in-depth study on the \Pattern{} patterns in the mobile ecosystem. Our research first highlights five distinct \Pattern{} patterns that compromise user experience, including \textit{text mislead}, \textit{UI mislead}, \textit{forced action}, \textit{out of context} and \textit{privacy-intrusive by default}. 
To further evaluate the impact of such \Pattern{} patterns at large, we developed an automated analysis pipeline called \system{}, to tackle the challenges of identifying, dismissing, and collecting diverse PoWs in real-world apps. Evaluation results showed that \system{} achieves high precision and recall in detecting PoWs, efficiently dismissed over 88\% of PoWs with minimal user interaction, with good robustness and reliability in comprehensive app exploration.
Further, our systematic analysis over the top 100 popular apps in China and U.S. revealing that both regions displayed significant ratios of \Pattern{} patterns, particularly in promotional contexts, with high occurrences in categories such as shopping and video apps. The findings highlight the strategic deployment of \Pattern{} tactics that compromise user trust and ethical app design.
\end{abstract}




\maketitle
\section{Introduction}
\label{sec:intro}



A pop-up window (PoW) in mobile apps is a small overlay that appears during user interactions, providing information or requesting actions to enhance user experience. However, excessive use of PoWs in modern apps has degraded usability. Many apps employ misleading text or UI elements to subtly manipulate decisions\cite{drakgrayhic2018,DP4}, force actions by limiting opt-out options\cite{autodetectdarkchen2023}, or display PoWs abruptly, thereby interrupting the user’s workflow. Additionally, some apps enable privacy settings by default, which can lead to inadvertent data disclosures \cite{privacyDefault1}. These tactics, aimed at boosting engagement and ad revenue, violate ethical design standards. We categorize these manipulative practices as \Pattern{} patterns.

\para{Challenges in Investigating the PoW Ecosystem}
Despite numerous frameworks proposed for automated app exploration, investigating the PoW ecosystem presents several inherent challenges:

\textbf{(1) PoW Identification.} Identifying PoWs is non-trivial due to their diverse types and intricate forms. PoWs can range from system permissions to custom functionalities or advertising prompts, often appearing in irregular shapes. This variability makes rule-based identification methods ineffective.

\textbf{(2) PoW Dismissal.} Dismissing PoWs is challenging due to the complex structure of their GUI component trees. These trees often include non-actionable background elements, leading to ineffective clicks if not precisely targeted. Additionally, PoWs embedded in mini-programs or WebViews are difficult to access directly, causing potential test failures.

\textbf{(3) Automated App Exploration} Automated traversal of commercial apps to collect PoWs faces significant hurdles. Effective state abstraction strategies are needed to avoid path explosion caused by frequent interface changes. Balancing exploration breadth and depth is crucial, as over-focusing on a single path limits coverage. Furthermore, modifying existing automated tests is difficult, as many tools provide only binary source code (e.g., JAR files\cite{AUIQtesting,AUIWtesting,AUIColumbus}).

\para{Our work} In this paper, we provide the first in-depth study on the \Pattern{} patterns in the mobile ecosystem. We have identified five distinct \Pattern{} patterns that compromise user experience: \textit{text mislead}, \textit{UI mislead}, \textit{forced action}, \textit{out of context} and \textit{privacy-intrusive by default}. Additionally, we categorize PoWs into promotional and functional types to provide a structured analysis of these manipulative practices. To further evaluate the impact of such \Pattern{} patterns at large, we have designed a automated app exploration tool \system{} to collect PoWs within an app.

To address the aforementioned challenges, we developed a set of novel mechanisms for automated PoW analysis:
\begin{itemize}
    \item \textbf{PoW Identifier} leverages computer vision-based object detection to identify diverse PoWs, while employing opacity analysis to reduce false positives through detailed analysis of GUI elements.
    \item \textbf{PoW Dismissal} interacts effectively with PoWs by identifying and prioritizing clickable components within them using a trained computer vision-based model. This strategy addresses the complex GUI component tree structure in PoWs, ensuring that interactions lead to the dismissal of PoWs without unnecessary clicks on non-actionable elements.
    \item \textbf{Automated App Exploration} employs strategies like Adaptive Depth-First Search and State Abstraction to enhance app exploration by adapting to frequent interface changes and minimizing path explosions. Additionally, Fault Tolerance and Path Recovery maintain an interface transition graph, allowing for smooth navigation back to previously visited interfaces.
\end{itemize} 

To comprehensively evaluate the effectiveness of \system{}, we conducted experiments focused on its three core modules: PoW Identifier, PoW Dismissal, and Automated App Exploration. First, using a dataset of 1,048 PoWs from various apps, our model achieved a precision of 98.3\%, recall of 94.4 \%, and an F1-score of 96.3\%, indicating high accuracy in identifying PoWs. Second, we tested \system{} on 50 apps, successfully dismissing 88\% of PoWs with no more than two clicks, demonstrating its efficiency in minimizing user interaction with intrusive interfaces. Lastly, we compared \system{}'s ability to collect PoWs with that of Monkey\cite{Monkey} and Q-Testing\cite{AUIQtesting}. \system{} outperformed both, showing \system{}'s robustness and reliability in app exploration.

\para{Measurement and findings} We analyzed the top 100 apps from China and the U.S., with a deeper focus on the top 20 Chinese apps across Shopping, Social, Video Streaming, and Tools \& Utilities categories. Our findings reveal that Chinese apps exhibited a higher frequency and a more complex array of PoW strategies compared to U.S. apps. 

Chinese apps, particularly in Shopping and Video Streaming, often use aggressive promotional and functional PoWs to drive user engagement. These tactics, often tied to time-sensitive promotions or essential notifications, may lead to impulsive decisions, raising concerns about user autonomy and ethical design practices. Furthermore, both Chinese and U.S. apps exhibit high ratios of \Pattern{} patterns, especially in promotional contexts, highlighting a widespread reliance on manipulative strategies to enhance engagement, potentially at the expense of user experience and trust.


In the spirit of open science, we will release the artifact of \system{}, as well as the corresponding dataset used in our research. \footnote{\url{https://github.com/feymanpaper/UIDarkPatternPopup}}.

\para{Contributions} We summarize our contributions as follows:
\begin{itemize}
    \item We provide the first study on privacy implications of \Pattern{} patterns in mobile apps and introduce key new understandings about the deceptive tactics employed by these PoWs.
    \item We introduce \system{}, an automated app exploration tool designed to detect, interact with, and collect PoWs in mobile applications. 
    \item We analyzed the top 100 apps from China and the U.S. to reveal the pervasiveness of \Pattern{} patterns. Besides, we found that \Pattern{} patterns are predominantly promotional across app categories such as shopping, video and social.
\end{itemize}


\section{Background and Motivation}
\label{sec:background}

\subsection{Pop-up Windows}

In modern applications, a Pop-up window (PoW) refers to a small overlay displayed prominently on the interface during user interactions. These PoWs provide information, or request actions, playing a key role in guiding user behavior and enhancing the user experience.
However, in many apps today, manipulative use of PoWs has degraded user experience. 
These tactics include misleading text or UI elements that subtly influence decisions, forced actions by hiding dismissal options, suddenly appearing PoWs that disrupt the user’s flow, and enabling privacy settings by default, leading to unintentional data sharing.
Such practices, aimed at boosting engagement and ad revenue, compromise usability and ethical standards. 

\subsection{\Pattern{} Patterns}
\label{subsec:back}


In light of the aforementioned issues with PoWs in contemporary applications, 
we categorize these tactics—\textit{Text Mislead, UI Mislead, Forced Action, Privacy-Intrusive by Default, Out of Context}—as \Pattern{} patterns.

\begin{figure*}
\centering
\subfloat[Text Mislead]{
\includegraphics[width=0.21\textwidth]{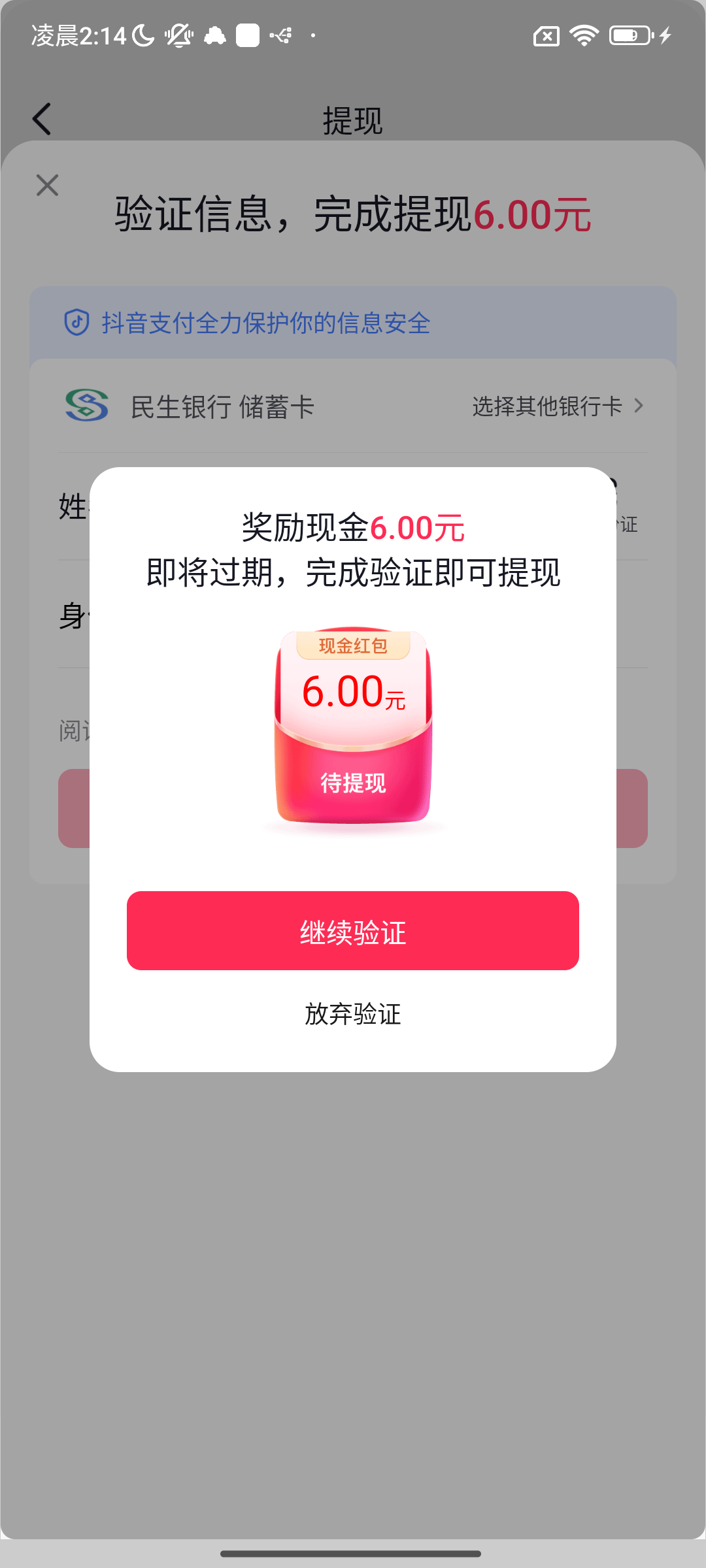} 
\label{fig:pattern_backmislead11}
} \hspace{6mm}
\subfloat[UI Mislead]{
\includegraphics[width=0.21\textwidth]{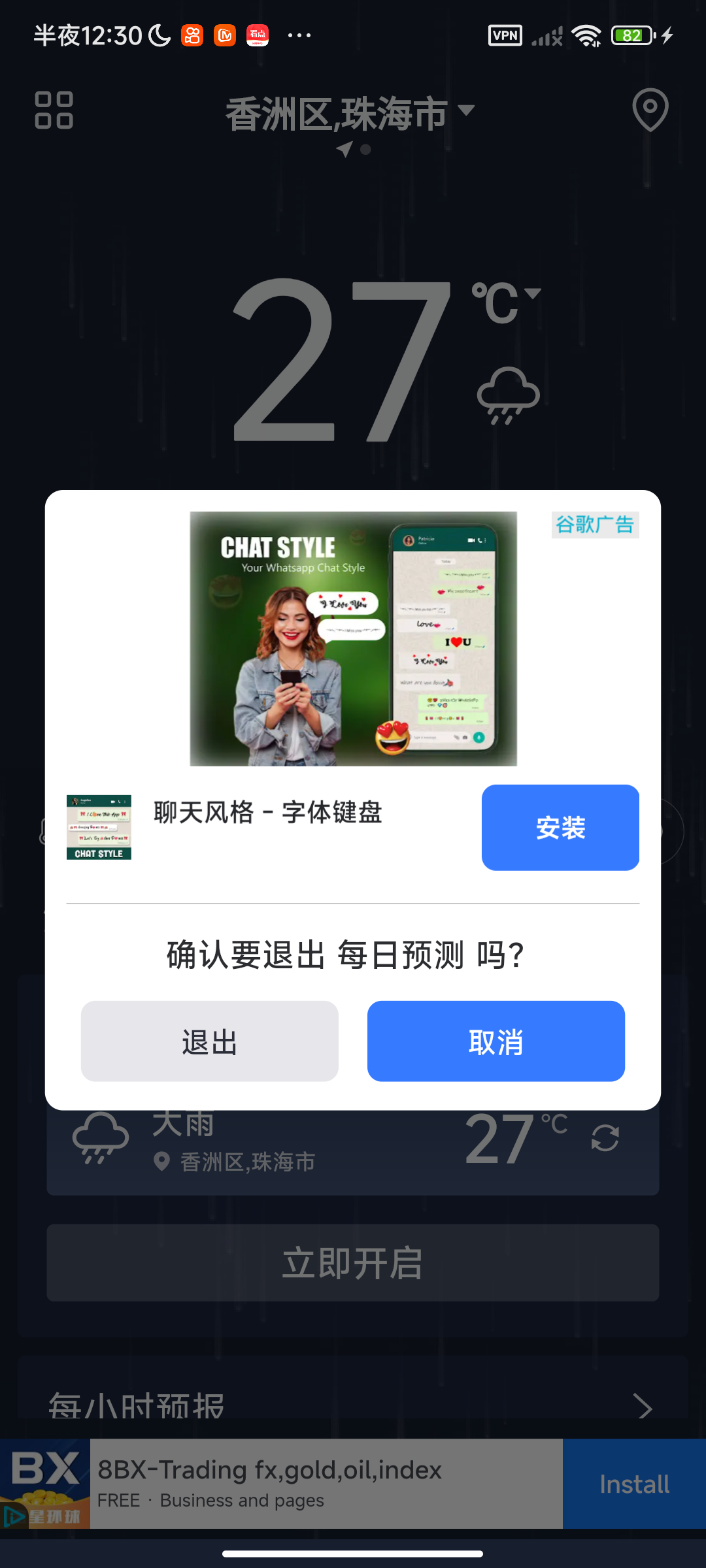} 
\label{fig:pattern_backmislead12}
} \hspace{6mm}
\subfloat[Forced Action]{
\includegraphics[width=0.21\textwidth]{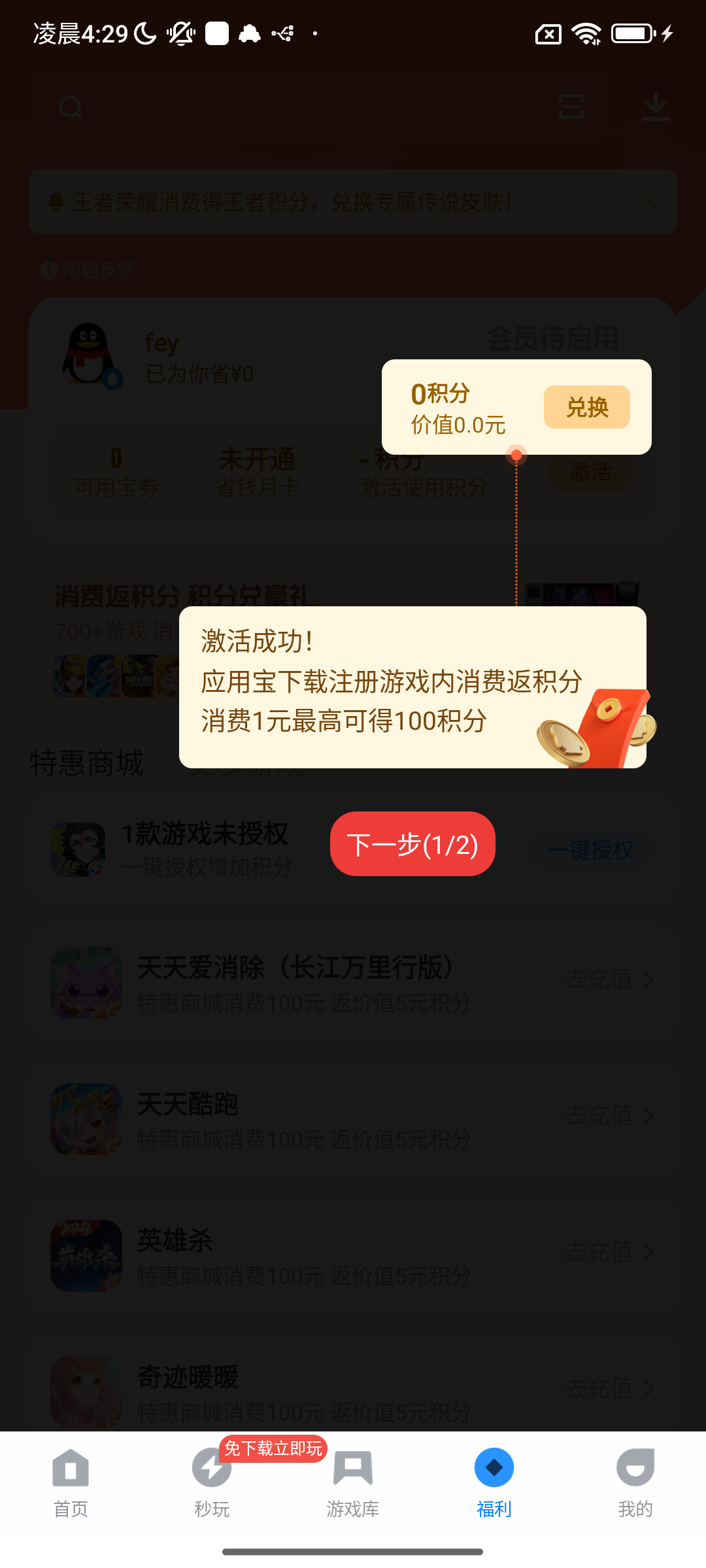}
\label{fig:pattern_noexitmislead1}
} \hspace{8mm}
\subfloat[Privacy-Intrusive by Default]{
\includegraphics[width=0.21\textwidth]{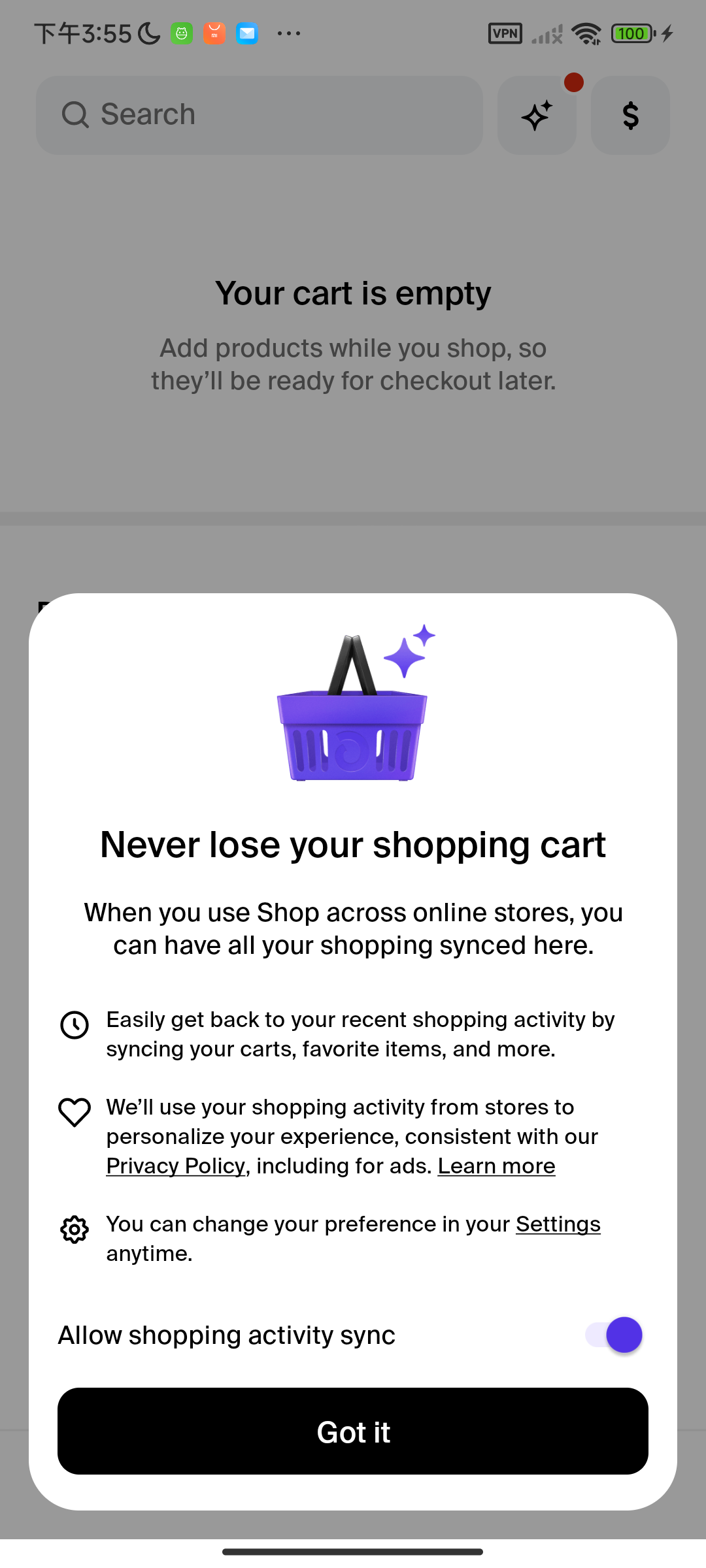}
\label{fig:pattern_privacydefaultnoexit1}
} \hspace{8mm}
\subfloat[Out of Context]{
\includegraphics[width=0.21\textwidth]{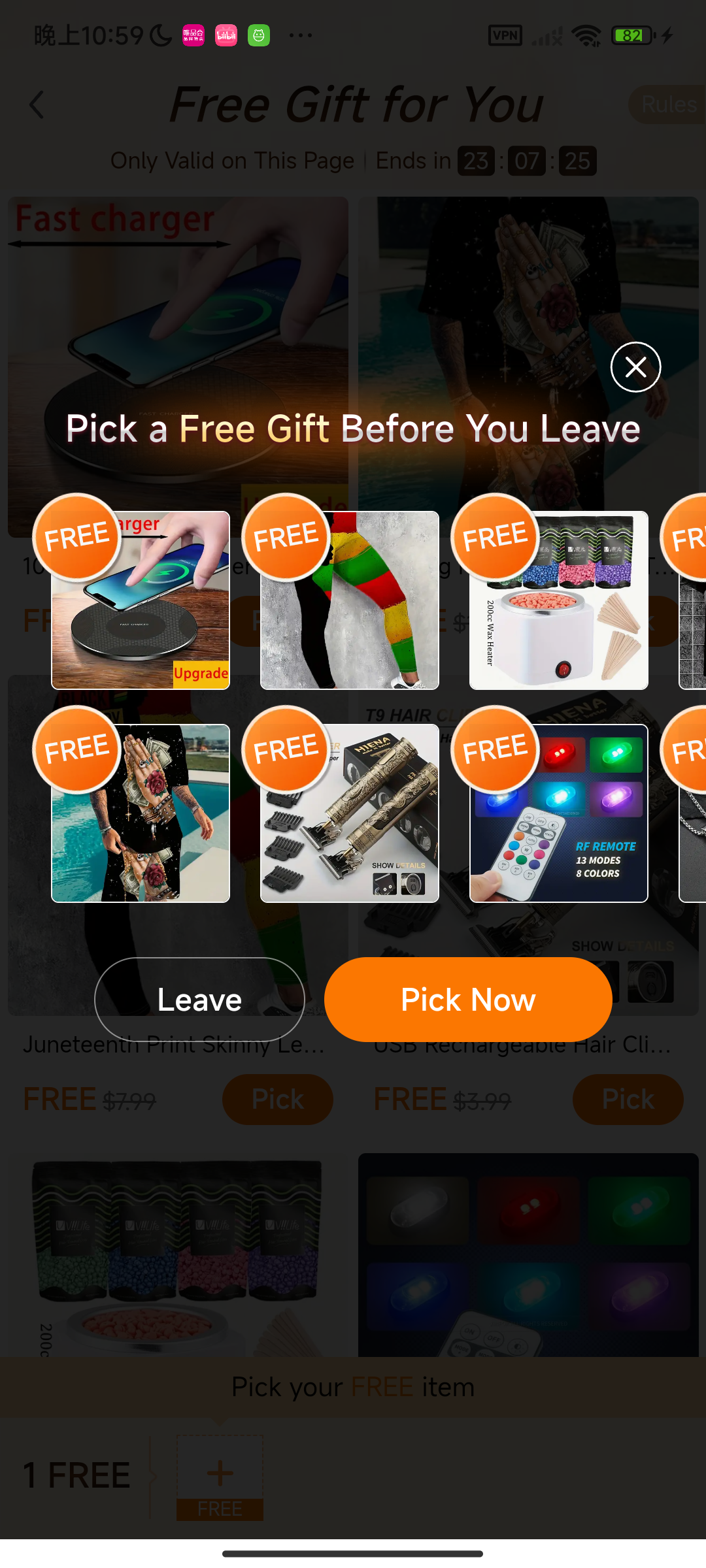}
\label{fig:pattern_outofcontext1}
} \hspace{6mm}  
\DeclareGraphicsExtensions.
\caption{Examples of \Pattern{} patterns in App PoWs discovered in our research.}
\label{fig:example-of-each-type-of-AUI}
\end{figure*}

\vspace{1pt}$\bullet$ \textbf{Text Mislead} This tactic involves the use of misleading or ambiguous language within PoWs to confuse users about the implications of their choices. Figure~\ref{fig:pattern_backmislead11} demonstrate how text can be crafted to mislead users into believing they are making a beneficial choice.

\vspace{1pt}$\bullet$ \textbf{UI Mislead} This tactic manipulates user behavior through interface design. Figure~\ref{fig:pattern_backmislead12} illustrates a PoW that appears within the "Confirm" and "Cancel" options, displaying an app download advertisement with an "Install" option. The "Install" button closely resembles the "Cancel" button, misleading users into mistakenly selecting the "Install" option.

\vspace{1pt}$\bullet$ \textbf{Forced Action} This tactic limits user choice by making it difficult to opt out. Figure~\ref{fig:pattern_noexitmislead1} depicts a PoW that appears when entering a new page, with a dark background highlighting a misleading offer of a small financial reward. With only a "Next" button available, this design forces users into decisions, frustrating and coercing them.

\vspace{1pt}$\bullet$ \textbf{Privacy-Intrusive by Default} This tactic involves pre-checking privacy settings in PoWs. Figure~\ref{fig:pattern_privacydefaultnoexit1} shows a PoW with pre-enabled settings, which can result in unintentional disclosure of personal information, contradicting user intent and violating informed consent principles.

\vspace{1pt}$\bullet$ \textbf{Out of Context} This tactic involves PoWs that appear at unexpected moments, disrupting the user’s natural flow. It includes PoWs that pop up suddenly or block users from navigating away. Figure~\ref{fig:pattern_outofcontext1} shows a PoW offering a free gift to entice users to stay when they attempt to leave the shopping interface, leading to unintended choices and undermining the principle of user control.

\section{Design of \system{}}
\label{sec:system}

\subsection{Workflow}

The workflow of \system{} involves a systematic process to identify, interact with, and analyze PoWs for potential \Pattern{} practices. Initially, the app under test is input into \system{}. \system{} then scans the current interface to detect the presence of a PoW. If a PoW is identified, \system{} focuses on dynamically testing all clickable components within the PoW to simulate user interactions and observe the PoW's behavior. Simultaneously, \system{} captures comprehensive context information about the PoW, including screenshots and GUI details. 
If no PoW is detected, \system{} systematically clicks through the app to trigger other potential PoWs within the same app.





\begin{figure}[htbp]
    \centering
    \includegraphics[width=0.7\linewidth]{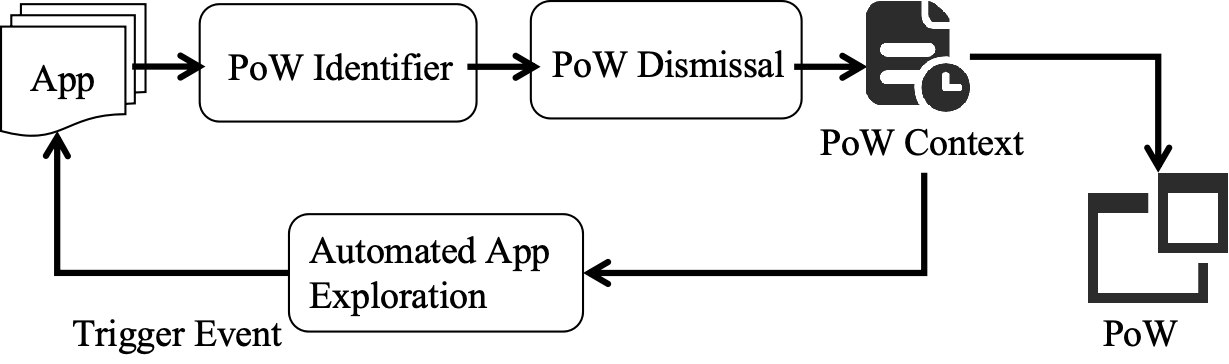}
    \vspace{-1em}
    \caption{Overview of \system{}.}
    \label{fig:fuzz}
\end{figure}

\subsection{PoW Identifier}

The PoW Identifier is designed to tackle the complex task of recognizing PoWs within the app. Recognizing these popups is non-trivial due to their diverse types and intricate forms. To effectively address this issue, we have adopted a computer vision-based object detection approach YOLO\cite{Yolo}. Based on empirical observation that PoW always overlays a semi-transparent gray mask on the background, we integrated an opacity analysis technique into the PoW Identifier to minimize false positives and enhance its reliability and accuracy. During testing, we use a predefined HSV (hue, saturation, value) interval to calculate the percentage of shadow in the image. After excluding the PoW area detected by YOLO, we determine the ratio of the shadowed region to the remaining image. If this ratio exceeds a set threshold, the image is classified as containing a PoW; otherwise, it's flagged as a false positive.

\subsection{PoW Dismissal}

The PoW Dismissal is specifically designed to address the challenge of effectively dismissing PoW by interacting with clickable components within them. This task is not straightforward due to the complex structure of the GUI component tree in PoW. Often, this GUI tree includes not only the clickable elements of PoW but also the background non-actionable elements counterparts, which do not contribute to dismissing the PoW. Direct interaction with these popups without precise targeting can lead to ineffective clicks on these background components, which do not facilitate the dismissal of the popup.
To resolve this issue, we have trained a YOLO model to identify the clickable elements within the PoWs, thereby avoiding interactions with non-actionable background components. However, considering the potential possibility of missed detections, we have implemented a fallback mechanism. This mechanism involves incorporating additional clickable components located within the popup into the interaction sequence as a secondary priority. These components are only interacted with if the primary clickable elements identified by YOLO are insufficient to dismiss the popup.

\subsection{Automated App Exploration}
Automated App Exploration aims to systematically collect PoWs during app traversal. It consists of three key modules: Adaptive Depth-First Search (ADFS), State Abstraction, and Fault Tolerance and Path Recovery. These modules work together to ensure efficient and comprehensive exploration of app interfaces.

\para{Adaptive Depth-First Search (ADFS)} Traditional depth-first search strategies often face challenges in balancing the exploration of shallow and deep interface hierarchies, especially in large and complex apps. The ADFS offers a robust and efficient approach for automated app interface exploration. By dynamically adjusting the traversal depth based on coverage metrics, ADFS ensures a balanced and thorough exploration of both shallow and deep interface hierarchies. 

    

\para{State Abstraction} For complex commercial apps, the number of pages is vast, and pages related to the same business often changes. Without state abstraction, we would traverse many similar interfaces repeatedly. Therefore, an appropriate state abstraction strategy is needed. We uniquely identify an interface using the positional information of all clickable components on the interface. Then, we employ the Longest Common Subsequence (LCS) algorithm \cite{LCSfast} to compare interface similarities and achieve state abstraction. Interface similarity comparison is necessary because the XML tree structure of interfaces, component texts, and component positions can all changes, and the LCS algorithm is more suitable for scenarios where subsequence positions have a related relationship.

\para{Fault Tolerance and Path Recovery} In ADFS strategy, we perform a rollback upon reaching the maximum traversal depth. Android maintains an activity stack for regular activities, allowing back operations to follow a FILO (First-In-Last-Out) approach. However, many app interfaces use fragments, which can complicate this. A single back operation may return directly to the main interface because Android sometimes does not store fragments in a stack, potentially destroying all stored fragments. To ensure we can return to previously traversed interfaces, we dynamically maintain the app's interface transition graph and restart the app upon exit, recovering the path based on this graph.

\section{Evaluation}
\label{sec:evaluation}



\para{Experimental Setup} 
Our experimental setup consists of three main components: Identifying PoWs, Dismissing PoWs, and App Exploration. For Identifying PoWs, we collected 1048 PoWs from various apps not included in the App Exploration dataset. For Dismissing PoWs, we utilized the AUI dataset\cite{cai2023darpa}, which comprises 1072 screenshots annotated with labels for App-guided Options and User-preferred Options. For App Exploration, we crawled the 50 most popular applications from Tencent's MyApp platform \cite{TencentMyApp2024}, covering categories like shopping, video, social, and utilities. We evaluated the effectiveness of Monkey, Q-Testing, and \system{} in detecting PoWs. To maintain consistency, each app was manually logged into, and all cache data was cleared prior to testing.

\subsection{Results and Analysis}

\para{Effectiveness of PoW Identification}
The performance metrics of the YOLO model for detecting PoW on the test set are as follows: Precision, Recall, and F1-score are 98.3\%, 94.4\%, and 96.3\%, respectively. An analysis of the false negatives and false positives revealed that false negatives were mainly found in instances where the color and boundaries of the PoW were not distinctly separable from the adjacent interface, such as in night mode and for PoWs with dark or grey fill colors. False positives were observed when loading the wait page or scanning the QR code. Aside from these rare occurrences, our model accurately identifies PoW. 

\begin{table}[htbp]
\centering
\caption{Effectiveness of \system{} in detecting PoWs} 
\label{tab:Effectiveness}
\begin{tabular}{cccc}
\hline 
\textbf{Type} & \textbf{Precision} & \textbf{Recall} & \textbf{F1-score} \\ 
\hline 
PoW & 0.983 & 0.944 & 0.963 \\ 
Confirmation Button & 1 & 0.949 & 0.976 \\ 
Exit Button & 1 & 0.917 & 0.965 \\
\hline 
\end{tabular}
\end{table}

\begin{figure}[ht]
\centering
\begin{subfigure}{0.56\textwidth}
\centering
\includegraphics[width=\linewidth]{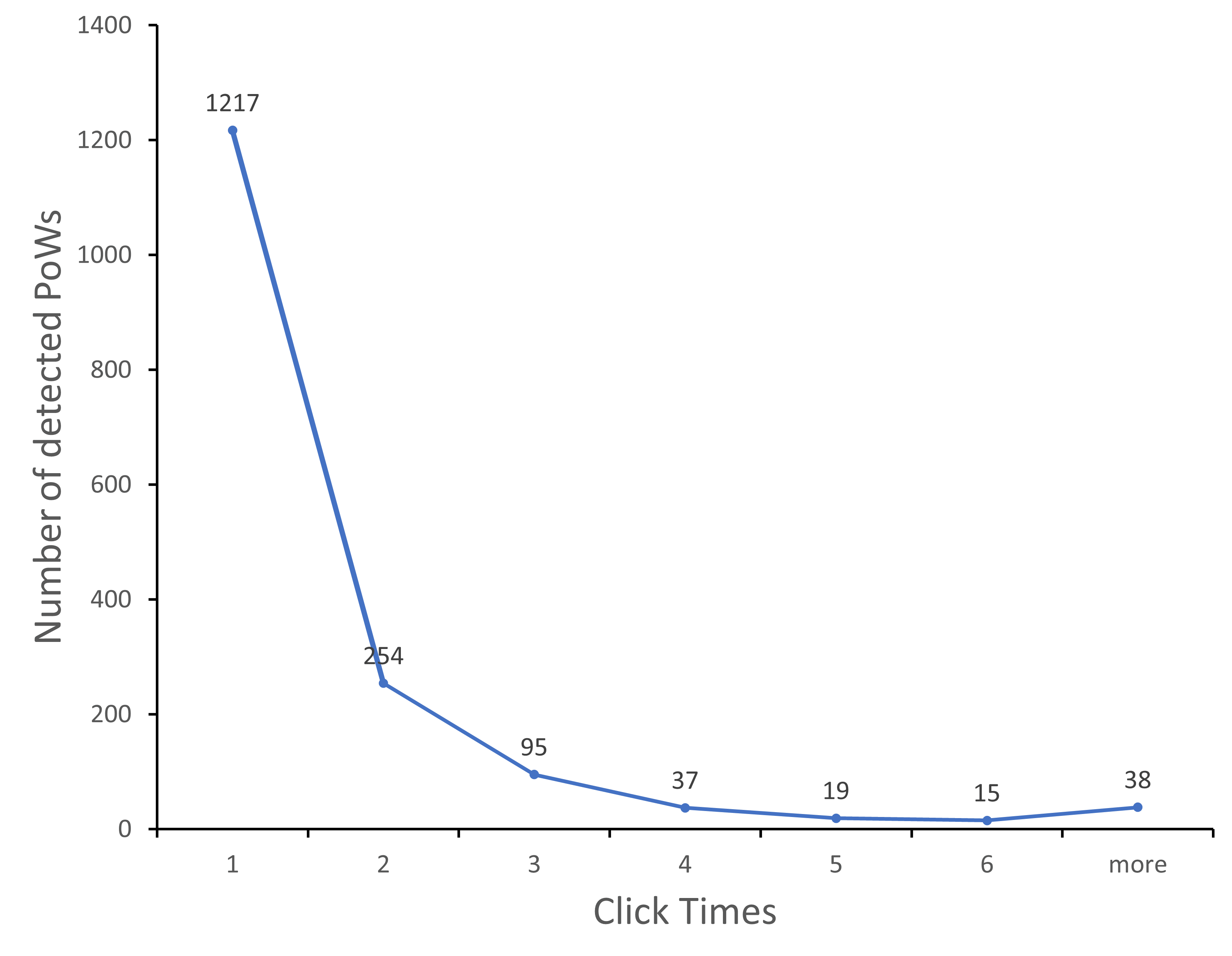} 
\caption{Number of clicks needed to dismiss PoWs in 50 apps}
\label{fig:powdismiss}
\end{subfigure}
\hfill 
\begin{subfigure}{0.40\textwidth}
\centering
\includegraphics[width=\linewidth]{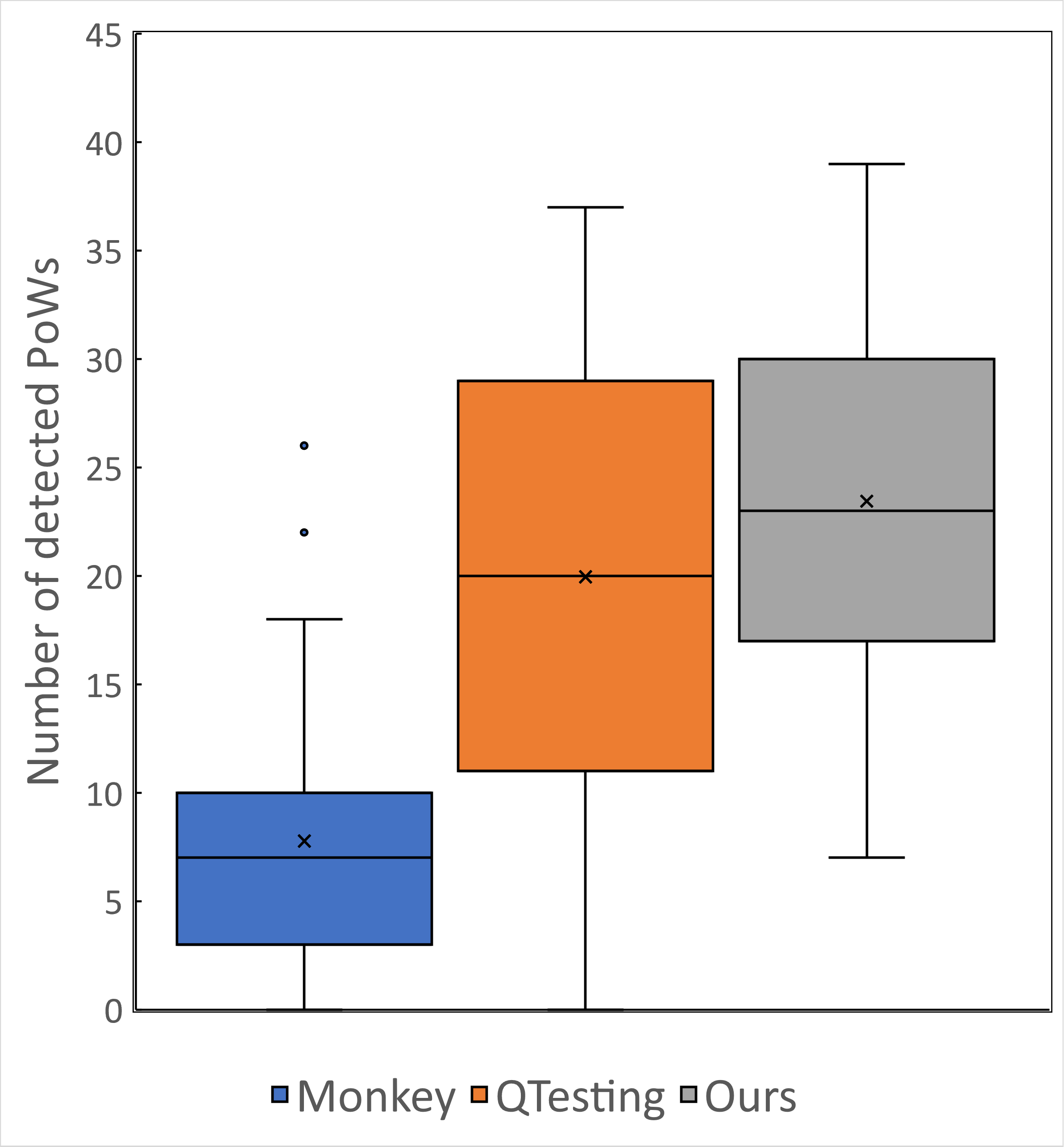} 
\caption{Effectiveness of different tools for detecting PoWs.}
\label{fig:compare_tool}
\end{subfigure}
\end{figure}

\para{Effectiveness in PoW dismissal} Figure~\ref{fig:powdismiss} reveals a steep decline in the number of PoWs dismissed as the number of clicks increases. Specifically, a substantial majority of PoWs, totaling 1,217, were dismissed with just one single click. This distribution underscores \system{}'s effectiveness in quickly resolving the majority of PoWs interruptions with minimal user interaction, as evidenced by the high number of PoWs dismissed with only one click. The ability to dismiss over 88\% of PoWs with no more than two clicks demonstrates \system{}’s capability in efficiently handling PoWs, which expedites the testing process.

The presence of a small proportion of PoWs requiring more than six clicks to dismiss can be attributed to two potential causes. First, the construction of the PoWs themselves might only include checkboxes or other non-dismissive interactive widgets. Second, the clickable components of the PoWs interface may encompass underlying clickable elements from the base interface. This overlap of interactive elements could necessitate multiple clicks to navigate through and finally dismiss the PoWs. 

\para{Effectiveness in App Exploration} Figure~\ref{fig:compare_tool} presents a comparative analysis of three automated exploration tool Monkey, Q-Testing, and \system{} in their ability to collect the number of PoW during the app exploration processes. 

From the box plots, it is evident that Monkey demonstrates the lowest median number of collected PoWs since Monkey is based on a random strategy. Besides, Q-Testing shows a lower median number of collected Pows compared to \system{}. Note that Q-Testing occasionally records zero PoWs collected in certain apps. Upon investigation, it appears that Q-Testing may be adversely affected by initial screen advertisements or frequent page transitions, such as video content. These elements seem to cause the tool to restart repeatedly. This issue highlights a potential vulnerability in Q-Testing's methodology, suggesting that its operational framework may not be robust against dynamic content changes, which are common in many modern app environments. Additionally, \system{} demonstrates a consistently higher collection capability of Pows, as evidenced by the highest median value and a comparably wide interquartile range, similar to that of Q-Testing. The absence of outliers in the data from our tool suggests a robust and stable performance across various testing scenarios.

\section{Understanding \Pattern{} Pop-Up Windows in the Wild}
\label{sec:measurement}

\para{Measurement setup} In our research, we systematically gathered data from the top 100 apps in China and the United States to assess the complexity of PoW ecosystems. Specifically, we selected the top 50 apps from the Tencent MyApp store\cite{TencentMyApp2024}, a prominent app market in China, and the top 50 apps from the Google Play Store\cite{googleplay} for the U.S. region. For each app, we conducted automated app exploration for one hour, during which we collected data on all PoWs encountered during the traversal process. Subsequently, we employed a semi-automated method to label the PoWs, facilitating a detailed analysis of their types and \Pattern{} patterns.




\subsection{\Pattern{} Patterns Identification}

We leverage LLMs, specifically ChatGLM \cite{GLM1}, to automate the detection of \Pattern{} patterns by analyzing textual and UI elements within apps. To enhance accuracy, we supplement the automated process with manual verification, ensuring precise classification and validation. This combination of automated detection and human validation ensures a reliable and comprehensive identification of sneaky patterns across mobile applications.

\para{Text Mislead} The primary indicator of a Text Mislead PoW is the use of misleading or ambiguous language that confuses users about the action they are taking. If the language subtly guides the user into performing an action that does not align with their initial intent, it is classified as Text Mislead.

\para{UI Mislead} This type involves a design or layout that visually misleads users into making unintended selections. If a PoW employs visual design elements to confuse the user into clicking on something that isn’t their intended action (such as making the "Cancel" button less noticeable), it’s classified as UI Mislead.

\para{Forced Action} This tactic forces users into a situation where they must interact with the PoW before proceeding, and there is no visible way to dismiss it. If the PoW provides no way to exit or bypass it, forcing the user to interact with it, it is classified as Forced Action.

\para{Privacy-Intrusive by Default} This tactic involves setting privacy-related options to the most permissive setting by default, often without proper user awareness. If the PoW enables certain privacy settings, like data collection by default, and the user is required to opt out instead of opting in, it is classified as Privacy-Intrusive by Default.

\para{Out of Context} These PoWs appear unexpectedly and disrupt the user’s flow, often at a critical or inopportune moment. For "Out of Context" categorization, we meticulously record the contextual relationships between interfaces during the automated testing process. Specifically, we track: (1). Interface transitions (e.g., clicks or back actions), (2). Textual content on the interface, and (3). Information about the clicked components. If no correlation is found between the PoW and the button click or the PoW is triggered by a back/return event, the PoW is classified as "Out of Context".


\subsection{PoW Type Classification}

We classify PoWs into two types: promotional PoW and functional Pow.

\vspace{1pt}\noindent$\bullet$ Firstly, we adopt pattern matching to identify functional PoWs requesting permissions (i.e., permission-related PoWs). We categorize the permission-related PoWs into two types. One is related to permissions at the Android system level, which can be distinguished based on the app package name in the XML information of the interface. The other type involves in-app permission request PoWs, which can be identified by fixed text patterns such as ``Allow \dots\ to access your location?'' or ``Enable \dots\ to send notifications.''

\vspace{1pt}\noindent$\bullet$  Then, we adopt LLM to identify if the remaining PoWs contain promotional information. Specifically, we use the Yolo model in the PoW identifier to obtain the bounding box of the PoW within a given UI screen. We then use OCR to extract all the texts it contains and send them to the LLM for promotional PoW identification.The prompt used for this task is fine-tuned to ensure the best performance.

\vspace{1pt}\noindent$\bullet$  Finally, the remaining PoWs that are not classified as promotional PoWs are classified as functional PoWs (permission-unrelated PoWs).

\subsection{Measurement results}
\subsubsection{Distribution, Ratio, and Count of Sneaky Patterns Across China and the U.S.}

Table~\ref{table:SDI_SBT_SDE3} provides a comprehensive analysis of the distribution, ratio, and count of \Pattern{} patterns across \PowProm{} and \PowFun{} categories in mobile applications within China and the United States. The data reveals significant regional differences not only in the distribution and ratio of these patterns but also in the sheer volume of PoWs and \Pattern{} instances, reflecting distinct strategic approaches to user engagement and consent management.

The total number of PoWs in Chinese apps is significantly higher than in U.S. apps, with 1,374 PoWs in China compared to 277 in the U.S. This stark difference suggests that Chinese apps are more aggressive in deploying PoWs, potentially due to a highly competitive market where user engagement and retention are critical. Furthermore, the number of \Pattern{} instances in China (808) is more than five times that in the U.S. (152), indicating a more pervasive use of \Pattern{} tactics in Chinese apps. This aligns with the observation that Chinese apps prioritize promotional strategies, as reflected in the high count of \PowProm{} PoWs (339) and \Pattern{} instances (323). In contrast, U.S. apps exhibit a more restrained approach, with fewer PoWs and \Pattern{} instances overall. This could be attributed to stricter regulatory environments and a greater emphasis on user privacy and transparency, which may limit the deployment of such tactics.

\noindent\fbox{%
\parbox{0.95\columnwidth}{%
\textbf{Finding 1: Chinese apps deploy significantly more PoWs and \Pattern{} instances than U.S. apps, with a pronounced focus on the \PowProm{} category, reflecting aggressive marketing strategies in a competitive market.}
}
}
\vspace{0.5em}

In China, \PowProm{} PoWs dominate the \Pattern{} category, accounting for 39.98\% of the total. This dominance is further emphasized by the high ratio of \Pattern{} instances within the \PowProm{} category (95.28\%). Such a high ratio suggests that Chinese apps heavily rely on \Pattern{} tactics in promotional contexts, likely to maximize user engagement and conversion rates. The \PowFun{} category also shows a significant presence of \Pattern{} patterns, particularly in the \PowPermiss{} subcategory (78.72\%), indicating a strategic use of these patterns to manage user permissions.

In the United States, the distribution of PoWs is more balanced, with \PowNecess{} PoWs constituting the majority within the \Pattern{} category at 52.63\%. The prevalence of \Pattern{} instances remains significant but is less pronounced compared to China, with \PowProm{} accounting for 90.32\%, \PowPermiss{} for 67.69\%, and \PowNecess{} for 44.20\%. This indicates that while U.S. applications also utilize \Pattern{} strategies, they tend to prioritize essential functionalities, likely influenced by stricter regulatory frameworks and a greater emphasis on obtaining user consent.

\noindent\fbox{%
\parbox{0.95\columnwidth}{%
\textbf{Finding 2: Both Chinese and U.S. apps exhibit high ratios of \Pattern{} patterns in Promotion, but the strategic deployment varies. Chinese apps focus more aggressively on promotional tactics.}
}
}
\vspace{0.5em}

\begin{table}[htbp]
\caption{Distribution and Ratio of \Pattern{} PoWs Across Two Categories (\PowProm{}, \PowFun{}).}
\small
\centering
\resizebox{0.8\linewidth}{!}{
\begin{tabular}{c|cccccc}
\hline
\multicolumn{1}{c}{\textbf{Region}} & \multicolumn{2}{c}{\textbf{Necessity}} & \begin{tabular}[c]{@{}c@{}}\textbf{\# PoWs} \\ \textbf{in App} \end{tabular} & \textbf{\# \Pattern} & \textbf{Distribution} & \textbf{Ratio} \\ 
\hline
\multirow{4}{*}{China} 
& \multicolumn{2}{c|}{\PowProm{}} & 339 & 323 & 39.98\% &95.28\% \\ \cline{2-3}
& \multirow{2}{*}{\PowFun{}}
    & \multicolumn{1}{|c|}{\PowPermiss{}} & 282 & 222 & 27.48\% &78.72\% \\
    && \multicolumn{1}{|c|}{\PowNecess{}} & 753 & 263 & 32.54\% &34.93\% \\ \cline{2-7}
& \multicolumn{2}{c|}{\textbf{Total}}   & 1374 & 808 & 100\% & \/ \\ 
\hline
\multirow{4}{*}{US} 
& \multicolumn{2}{c|}{\PowProm{}}  & 31 & 28 & 18.42\% &90.32\% \\ \cline{2-3}
& \multirow{2}{*}{\PowFun{}} 
    & \multicolumn{1}{|c|}{\PowPermiss{}}  & 65 & 44 & 29.95\% &67.69\% \\
    && \multicolumn{1}{|c|}{\PowNecess{}} & 181 & 80 & 52.63\% &44.20\% \\ \cline{2-7}
& \multicolumn{2}{c|}{\textbf{Total}}  & 277 & 152 & 100\% & \/ \\ 
\hline
\end{tabular}
}
\label{table:SDI_SBT_SDE3}
\end{table}

\subsubsection{Prevalence of Sneaky Patterns in Popular Apps Across China and the U.S.}

Tables~\ref{table:top15_cn} and~\ref{table:top15-usa} present an analysis of mobile applications in China and the United States, respectively, focusing on the prevalence of \Pattern{} patterns within these apps. The tables categorize the top apps by their ratio of \Pattern{} patterns to total PoWs, providing insights into the extent of potentially intrusive or deceptive practices employed by these apps. 

\begin{table}[htbp]
\caption{\centering Top 10 apps with \Pattern{} PoWs in China.}
\centering
\resizebox{0.8\linewidth}{!}{
\begin{tabular}{ccccccc}
\hline
\textbf{App Name} & \textbf{Package Name}                           & \textbf{Version }                              & \textbf{Downloads} & \textbf{Category} &\textbf{Sneaky Count} & \textbf{Ratio}       \\ \hline
Mango TV  & com.hunantv.imgo.activity               & 8.1.4                                & 366M+    & Video \&Audio      &23            &92.00\% \\
Qimao novel  & com.kmxs.reader                  & 7.48                                & 47M+    & News \& Magazines  &32                            &84.21\%           \\
Kwai  & com.smile.gifmaker                     & 12.4.30.36528 & 2.4B+    & Video \& Audio           &21                  &80.77\% \\
Ctrip  & ctrip.android.view                      & 8.69.6                              & 376M+     & Travel \& Local     & 21                        &77.78\% \\
KuGou Music & com.kugou.android                        & 12.2.8                              & 2.9B+     & Video \& Audio      &45                &76.27\% \\
Xigua  & com.ss.android.article.video                        & 8.5.4                               & 953M+     & Video \& Audio      &25               &75.76\% \\
Himalaya & com.ximalaya.ting.android                     & 9.2.51.3                                & 427M+     & Video \& Audio        &37     &75.51\% \\
YOUKU  & com.youku.phone & 11.0.82        & 2.1B+     & Video streaming     &17                         &73.91\% \\
Toutiao Search  & com.ss.android.article.lite                       & 9.8.1.0        & 678M+     & Tools \& Utilities            &28      &73.68\% \\
WeSing & com.tencent.karaoke                         & 8.21.138.278                               & 956M+     & Video \& Audio   &33   &71.74\% \\
 \hline
\end{tabular}
}
\label{table:top15_cn}
\end{table}

\begin{table}[ht]
\caption{\centering Top 10 Apps with \Pattern{} PoWs in U.S.}
\centering
\resizebox{0.8\linewidth}{!}{
\begin{tabular}{ccccccc}
\hline
\multicolumn{1}{l}{\textbf{App Name}} & \textbf{Package Name}            & \textbf{Version} & \textbf{Downloads} & \textbf{Category} &\textbf{Sneaky Count}& \textbf{Ratio}       \\ \hline
PDF reader                      & com.ezt.pdfreader.pdfviewer            & 8.1.4  & 100M+     & Tools \& Utilities &9  &100\%       \\
CapCut                      & com.lemon.lvoverseas        & 11.9.0 & 1B+     & Video \& Audio   &5 &83.33\% \\
Shop                      & com.shopify.arrive        & 2.157.0 & 41M+      & Shopping  & 8 &72.73\%  \\
Instagram                      & com.instagram.android & 332.0.0.38.90  & 5.6B+       & Social networking &6 &66.67\%  \\
Aliexpress                      & com.alibaba.aliexpresshd            & 8.98.9 & 690M+       & Shopping &7 & 63.64\%         \\ 
NewsBreak                      & com.particlenews.newsbreak            & 24.21.0  & 76M+     & News \& Magazines &5  & 62.50\%         \\
Temu                      & com.einnovation.temu        & 2.68.0 & 320M+    & Shopping &6 &60.00\%  \\
LinkedIn                      & com.linkedin.android        & 4.1.942 & 1.6B+      & Tools \& Utilities &11   &57.89\%  \\
Bigo Live                      & sg.bigo.live & 6.13.1  & 660M+       & Social networking  &16     &57.14\%  \\
Canva                      & com.canva.editor            & 2.264.0 & 420M+       & Tools \& Utilities   & 7    &43.75\%    \\
\hline
\end{tabular}
}
\label{table:top15-usa}
\end{table}

The data from both tables reveals a significant reliance on \Pattern{} patterns in apps with large user bases, as indicated by the high download numbers. This could reflect a broader trend where developers use such tactics to capture attention or drive certain behaviors, which might not always align with user expectations or preferences.

Furthermore, the presence of \Pattern{} patterns in essential app categories such as video and utilities applications suggests that these tactics are not limited to more casual or entertainment-oriented apps but are also prevalent in apps used for productivity and information consumption. This could raise concerns about user experience, trust, and the ethical implications of app design practices.

\vspace{0.5em}
\noindent\fbox{%
\parbox{0.95\columnwidth}{%
\textbf{Finding 3: Popular apps in both China and the U.S. have significant proportions of \Pattern{} patterns, particularly prevalent in video, shopping, social, and utilities.}
}
}
\vspace{0.5em}

\subsubsection{Distribution, Ratio of Sneaky Patterns in Shopping, Video, Social, and Tools \& Utilities Apps}

\begin{figure}[htbp]
    \centering
    \includegraphics[width=0.8\linewidth]{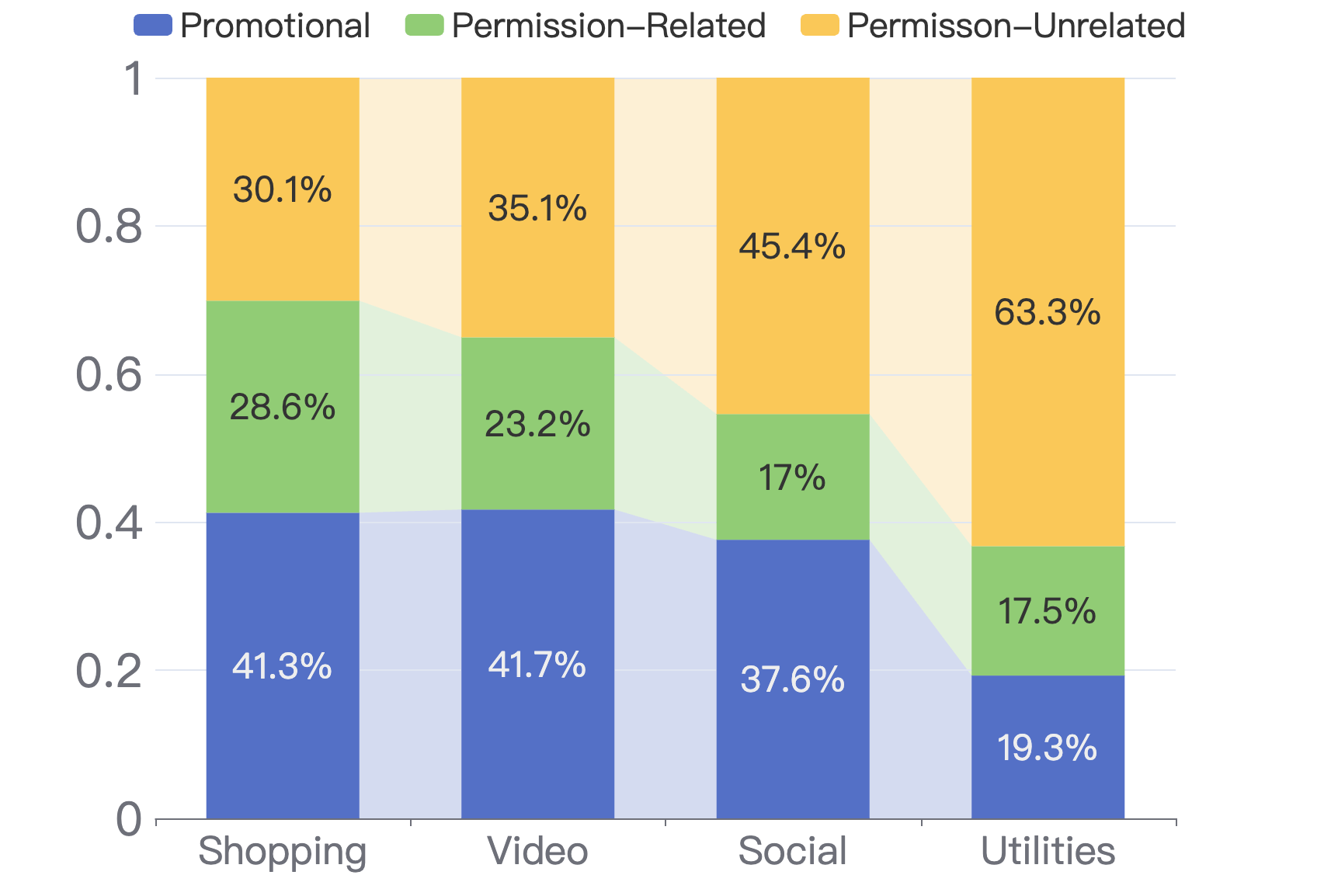}
    \vspace{-1em}
    \caption{Distribution of \Pattern{} patterns Across Two Categories(\PowProm{}, \PowFun{}) in Shopping, Video, Social, and Tools \& Utilities Apps}
    \label{fig:fuzz6}
\end{figure}

Figure~\ref{fig:fuzz6} shows the distribution of \Pattern{} patterns across app categories reveals distinct trends. Promotional patterns (blue bars) are most prevalent in shopping (41.3\%) and video (41.7\%) apps, indicating a strong focus on marketing and user engagement in these categories. Functional-permission-unrelated patterns (orange bars) dominate in utilities apps (63.3\%), suggesting that these apps frequently request business logic interactions to enhance functionality.

\vspace{0.5em}
\noindent\fbox{%
\parbox{0.95\columnwidth}{%
\textbf{Finding 4: The distribution of \Pattern{} patterns reveals that promotional patterns are most prevalent in shopping and video apps, highlighting a strong emphasis on marketing and user engagement in these categories.}
}
}
\vspace{0.5em}



The table~\ref{table:table2} highlights the prevalence of \Pattern{} patterns across different app categories, with a focus on promotional (\PowProm{}) and functional (\PowFun{}) patterns. Promotional patterns exhibit a high sneakiness ratio across all categories, particularly in video (94.17\%) and social (93.18\%) apps, indicating a strong tendency for these apps to employ marketing strategies that may be perceived as intrusive. Functional patterns related to permissions (\PowPermiss{}) also show significant sneakiness, especially in video (78.75\%) and social (80.43\%) apps, suggesting that these apps often tend to induce users to grant permissions. This data underscores the varying degrees of user experience concerns across app categories, with promotional and permission-related patterns being the most prevalent and potentially problematic.



\vspace{0.5em}
\noindent\fbox{%
\parbox{0.95\columnwidth}{%
\textbf{Finding 5: Promotional patterns are highly prevalent across all app categories, with the highest sneakiness ratios in video and social apps.}
}
}
\vspace{0.5em}

\begin{table}[htbp]
\centering
\caption{Ratio of \Pattern{} Patterns Across Two Categories (\PowProm{}, \PowFun{}) in Shopping, Video, Social, and Tools \& Utilities App}
\small
\resizebox{0.8\linewidth}{!}{
\begin{tabular}{c|ccccc}
\hline
\multicolumn{1}{c}{\textbf{Category}} & \multicolumn{2}{c}{\textbf{Type}} & \begin{tabular}[c]{@{}c@{}}\textbf{\# Pows} \\ \textbf{in App} \end{tabular} & \textbf{\# \Pattern} & \textbf{Ratio} \\ 
\hline
\multirow{3}{*}{Shopping} 
& \multicolumn{2}{c|}{\PowProm{}} & 120 & 111 & 92.50\% \\ \cline{2-3}
& \multirow{2}{*}{\PowFun{}}
    & \multicolumn{1}{|c|}{\PowPermiss{}} & 103 & 77 & 74.76\% \\
    && \multicolumn{1}{|c|}{\PowNecess{}} & 262 & 81 & 30.92\% \\ \cline{2-6}
\hline
\multirow{3}{*}{Video} 
& \multicolumn{2}{c|}{\PowProm{}}  & 120 & 113 & 94.17\% \\ \cline{2-3}
& \multirow{2}{*}{\PowFun{}}
    & \multicolumn{1}{|c|}{\PowPermiss{}}  & 80 & 63 & 78.75\% \\
    && \multicolumn{1}{|c|}{\PowNecess{}} & 195 & 95 & 48.72\% \\ \cline{2-6}
\hline
\multirow{3}{*}{Tools \& Utilities} 
& \multicolumn{2}{c|}{\PowProm{}}  & 40 & 32 & 80.0\% \\ \cline{2-3}
& \multirow{2}{*}{\PowFun{}} 
    & \multicolumn{1}{|c|}{\PowPermiss{}}  & 39 & 29 & 74.36\% \\
    && \multicolumn{1}{|c|}{\PowNecess{}} & 273 & 105 & 38.46\% \\ \cline{2-6}
\hline
\multirow{3}{*}{Social} 
& \multicolumn{2}{c|}{\PowProm{}}  & 88 & 82 & 93.18\% \\ \cline{2-3}
& \multirow{2}{*}{\PowFun{}} 
    & \multicolumn{1}{|c|}{\PowPermiss{}}  & 46 & 37 & 80.43\% \\
    && \multicolumn{1}{|c|}{\PowNecess{}} & 224 & 99 & 44.20\% \\ \cline{2-6}
\hline
\end{tabular}
}
\label{table:table2}
\end{table}

\subsection{Case Studies} 
\begin{figure}
\centering
\subfloat[Case 1]{
\includegraphics[width=0.20\textwidth]{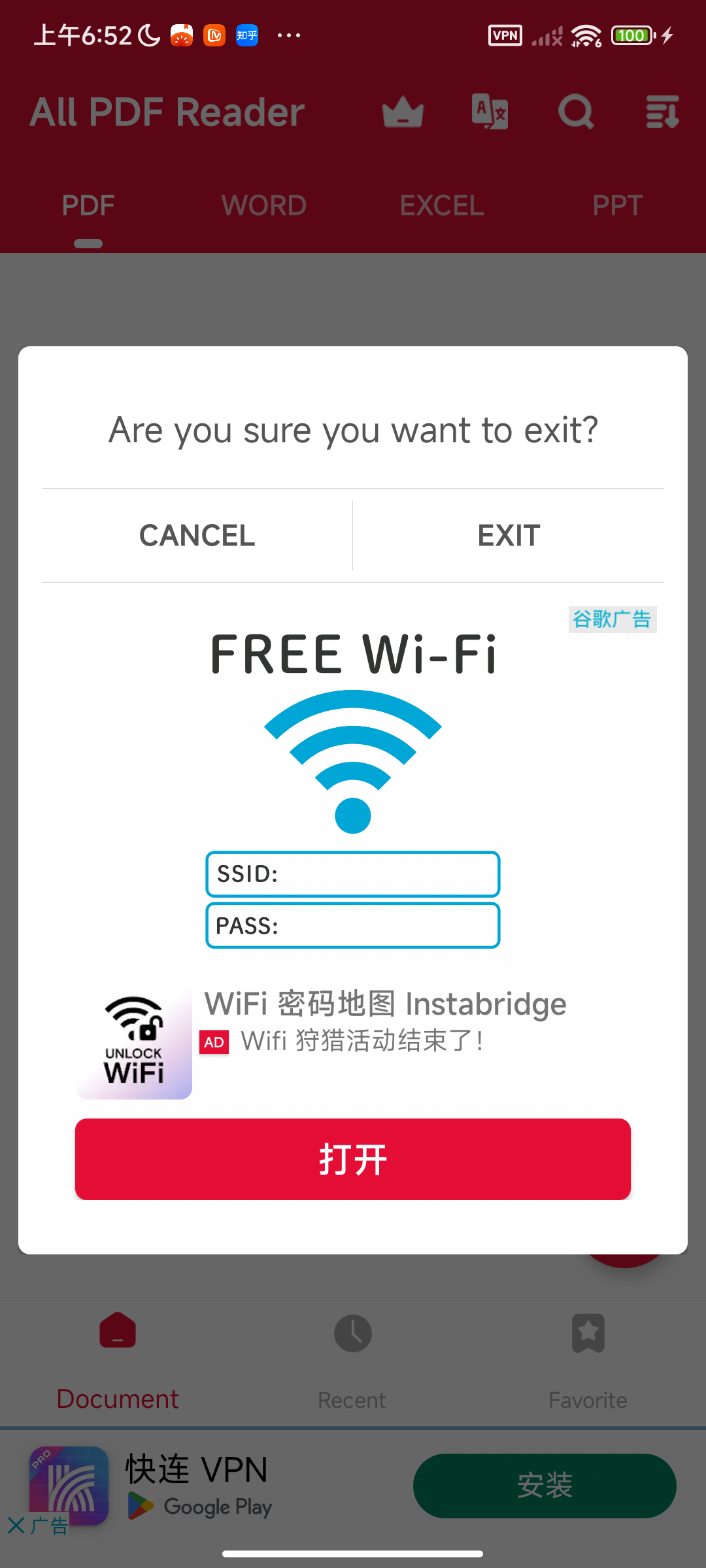} 
\label{fig:pattern_backmislead}
} \hspace{5mm}
\subfloat[Case 2]{
\includegraphics[width=0.20\textwidth]{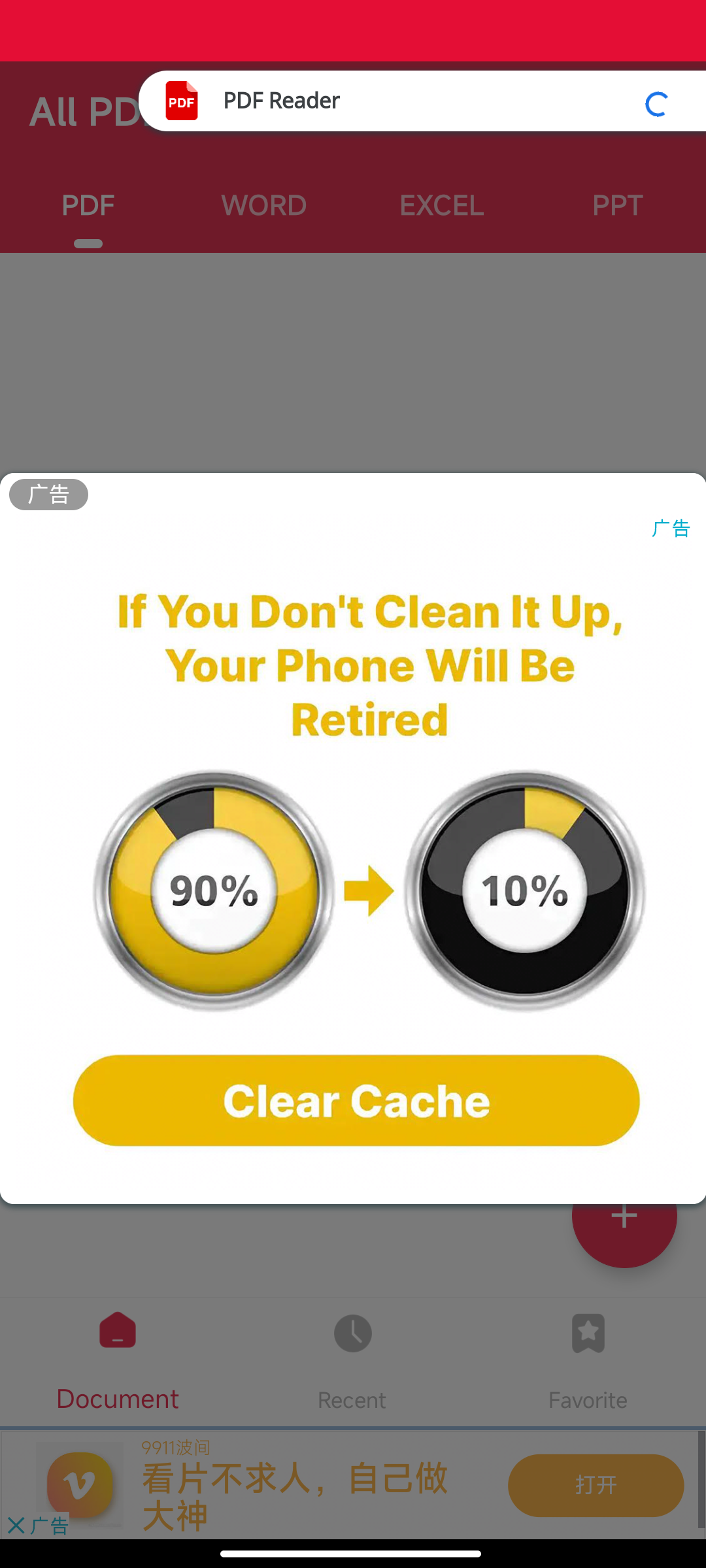}
\label{fig:pattern_privacydefaultnoexit}
} \hspace{5mm}
\DeclareGraphicsExtensions.
\caption{\Pattern{} Patterns in "All PDF Reader" App}
\label{fig:example-of-each-type-of-AUI 2}
\end{figure}

We present case studies to discuss the identified \Pattern{} patterns in a single App. Figure~\ref{fig:pattern_backmislead} shows how ads hijack the back button when users attempt to exit the app, using misleading text and deceptive buttons that seamlessly blend into the interface, creating confusion. Additionally, as seen in Figure~\ref{fig:pattern_privacydefaultnoexit}, PoWs appear unexpectedly upon app entry, without any user interaction. The ad features fake system warnings and deceptive personal notifications, exploiting user trust. The \Pattern{} advertising tactics in "All PDF Reader" raise serious ethical concerns, degrading user experience and potentially compromising security by promoting the download of unwanted software.

\section{Discussion and Limitation}
\label{sec:discuss}
\para{Extending Poker to iOS} Our research only provides a lower bound to evaluate the severity of \Pattern{} patterns. Poker currently does not cover iOS apps due to the fundamental differences in automated app analysis setup (i.e., UIAutomator\cite{UIAutomator} only supports Android), which is out of the scope of this research. In practice, Poker can be easily migrated to iOS by integrating the corresponding app automation framework (e.g., Appium\cite{appium} and WDA\cite{WDA}).

Besides, our preliminary analysis showed that the prevalence of PoWs in iOS is similar to that in the Android ecosystem. This is because neither of the two ecosystems provides any regulation or limitation regarding PoWs.

\para{Semi-Automated Sneaky Pattern Identification}
To ensure the accuracy of our classification results, we employed a semi-automated method to categorize each PoW and identify its \Pattern{} patterns. It is important to note that due to the complexity of \Pattern{} patterns, current methods, including multimodal large models, are unable to accurately classify them. In future work, we aim to develop a fully automated approach to enhance the efficiency, accuracy, and scalability of the identification process.

\section{Related Work}
\para{Consent Manipulation and Data Collection}
Prior research has explored UI dark patterns in consent manipulation, excessive data collection, and unintended user actions. Studies on permission requests~\cite{cao2021large} show how design elements like dismissal mechanisms and financial incentives influence user behavior. GDPR-related research on consent management platforms~\cite{Related_cookies1,Related_cookies2,ConsentNouMidas,DP4} reveals systematic use of dark patterns in web-based cookie banners. Investigations~\cite{relatred_and1,bramboon2017} highlight inadequate consent practices affecting millions of mobile users. However, these studies focus mainly on permissions and notifications, leaving a gap in understanding the manipulative potential of PoWs in mobile ecosystems. Poker addresses this gap by providing the first systematic analysis of PoWs for sneaky practices.


\para{Dark Patterns in Mobile Apps} Recent studies highlight the prevalence of dark patterns in mobile ecosystems. The research~\cite{Related_user2,Related_user1} shows that 95\% of apps employ dark patterns, and despite users recognizing these patterns, they often fail to fully understand their potential harms. Gunawan et al.~\cite{ComparativeStudyMMGuan} found mobile apps employ more manipulative designs than web platforms. This observation aligns with and substantiates our findings regarding the prevalence of dark patterns in mobile ecosystems. Long et al.~\cite{LongDarkPattern} systematically analyzed dark patterns in Chinese mobile apps, focusing on deceptive tactics like disguised ads and misleading notifications. However, their work does not specifically address PoWs as a distinct category of manipulative design. Poker builds upon these insights by innovatively combining existing techniques for app exploration (e.g., UIAutomator), PoW identification and dismissal (e.g., YOLO), and PoW categorization (e.g., LLMs) to systematize the manifestation of dark patterns in PoWs.


\para{Automated App Exploration} Prior research has developed techniques to enhance code coverage and detect bugs causing app crashes. Automated exploration methods fall into three categories: 1.Model-based approaches~\cite{AUITimeTravel} abstract the app and use exploration algorithms to generate test cases. 2.Dynamic and static analysis~\cite{AUIColumbus,AUIPermdroid,AUIWtesting} combines runtime behavior with APK or code analysis to guide UI traversal. 3.Learning-based methods~\cite{AUIQtesting,AUIGPT3Testing} employ machine learning and LLMs to identify patterns, improving scenario detection in future tests. Poker extends these approaches by introducing novel strategies for PoW identification and dismissal, addressing gaps in prior research and enabling scalable exploration of the PoW ecosystem.

\section{Conclusion}
\label{sec:conclusion}
In this paper, we conducted an in-depth analysis of \Pattern{} patterns in mobile apps, identifying five distinct patterns that compromise user experience: \textit{text mislead}, \textit{UI mislead}, \textit{forced action}, \textit{out of context} and \textit{privacy-intrusive by default}. To highlight the prevalence of these issues, we developed an automated app exploration tool named \system{}, which effectively detects and interacts with PoWs during automated app exploration. Our findings indicate that Chinese apps demonstrate a higher frequency of \Pattern{} patterns, particularly in promotional contexts, with significant occurrences in the shopping and video domains. These results underscore the need for greater awareness of ethical app design and the potential impact of \Pattern{} patterns on user trust and experience.



\bibliographystyle{plain}
\bibliography{refs}

\end{document}